# Anisotropic Infrared Response and Orientation-dependent Strain-tuning of the Electronic Structure in Nb$_2$SiTe$_4$

*Fanjie Wang [†][#], Yonggang Xu [‡][#], Lei Mu [†], Jiasheng Zhang [†], Wei Xia [§], Jiamin Xue [§], Yanfeng Guo [§][∥] *, Ji-Hui Yang [‡]*, and Hugen Yan [†]**

Fanjie Wang and Yonggang Xu contributed equally to this work.

[†] State Key Laboratory of Surface Physics, Key Laboratory of Micro and Nano-Photonic Structures (Ministry of Education) Department of Physics, Fudan University, Shanghai 200433, China.

[‡] Key Laboratory for Computational Physical Sciences (MOE), State Key Laboratory of Surface Physics, Department of Physics, Fudan University, Shanghai 200433, China
Shanghai Qizhi Institution, Shanghai 200232, China

[§] School of Physical Science and Technology, ShanghaiTech University, Shanghai 201210, China.

[∥] ShanghaiTech Laboratory for Topological Physics, ShanghaiTech University, Shanghai 201210, China

*E-mail: guoyf@shanghaitech.edu.cn (Y. Guo)

jhyang04@fudan.edu.cn (J.-H. Yang)

hgyan@fudan.edu.cn (H. Yan)

**ABSTRACT**

Two-dimensional materials with tunable in-plane anisotropic infrared response promise versatile applications in polarized photodetectors and field-effect transistors. Black phosphorus is a prominent example. However, it suffers from poor ambient stability. Here, we report the strain-tunable anisotropic infrared response of a layered material $Nb_2SiTe_4$, whose lattice structure is similar to the 2H-phase transition metal dichalcogenides (TMDCs) with three different kinds of building units. Strikingly, some of the strain-tunable optical transitions are crystallographic axis-dependent, even showing opposite shift when uniaxial strain is applied along two in-plane principal axes. Moreover, $G_0W_0$-BSE calculations show good agreement with the anisotropic extinction spectra. The optical selection rules are obtained via group theory analysis, and the strain induced unusual shift trends are well explained by the orbital coupling analysis. Our comprehensive study suggests that $Nb_2SiTe_4$ is a good candidate for tunable polarization-sensitive optoelectronic devices.

In-plane anisotropy of two-dimensional materials is of great importance for directional transport of charge and energy. One well-known example is black phosphorous (BP). BP has a layer-dependent bandgap from 0.3 to 2.0 eV,[1-3] and has been shown to host highly in-plane anisotropic excitons[4-7] and theoretically predicted to exhibit elliptic and hyperbolic dispersion for polaritons.[8-11] Although few-layer BP is promising in infrared photodetectors and modulators, it is very unstable in the ambient condition.[12-14] On the other hand, traditional 2H-phase transition metal dichalcogenides (2H-TMDCs)[15] are not suitable for infrared photonic applications because the bandgap is typically in the visible range. Thus, seeking a stable anisotropic two-dimensional (2D) material with infrared bandgap is highly desirable.

$Nb_2SiTe_4$ (NST) is a layered material, composed of stacked Te-($NbSi_{0.5}$)-Te sandwich layers, which can be considered as a result of intercalation of Si into the 2H-TMDCs. It exhibits similar structural properties to the 2H-TMDCs with three different kinds of building units. NST was synthesized in 1994 by W. Tremel et al..[16] Previous theoretical calculations suggest that monolayer NST is an indirect-gap semiconductor with a band gap of 0.84 eV and the bulk counterpart exhibits a reduced bandgap of 0.39 eV. Experimentally, scanning tunneling spectroscopy (STS) gives a bandgap of 0.39 eV for the bulk.[17] Field-effect transistors based on few-layer NST show relatively high carrier mobility of ~100 $cm^2V^{-1}s^{-1}$. The anisotropic transport,[18-21] excellent ferro-elasticity,[22] and anisotropic thermoelectric properties[23] are also theoretically predicted. By changing its elemental compositions, the resultant materials also attract tremendous attention. These materials include other phases with different contents of Si and, in particular, metallic Nb$_3$SiTe$_6$.[24-25] An unexpected enhancement of the weak-antilocalization signature in magneto-transport was experimentally observed in few-layer Nb$_3$SiTe$_6$.[24] Moreover, nonsymmorphic-symmetry-protected hourglass Dirac loop, nodal line, as well as Dirac point were predicted in bulk and monolayer Nb$_3$SiTe$_6$.[25] Narrow-gap semiconducting NST has a suitable bandgap and exhibits good environmental stability, which make it a competent candidate for infrared optoelectronic devices.[26] However, the optical inter-band transitions of NST have never been investigated experimentally, and the anisotropic optical responses in the infrared range are essentially unknown, let alone the strain-tunability of the optical transitions.

Here, we show that NST has a series of polarization-dependent transition resonances and a linear dichroism switch, which reveals the underlying electronic structure and gives us a reliable way to determine the crystallographic orientation. More interestingly, the strain-induced band engineering is orientation-dependent. Some of the transitions exhibit anomalous blueshift with the tensile strain along one of the crystallographic axes. Our theoretical calculations and group theory analysis well explain the anisotropic optical resonances. Furthermore, the observed orientation-dependent strain engineering is fully captured by the orbital coupling analysis. Our work broadens the in-depth understanding of the optical properties of NST, and establishes NST as a good candidate for tunable optoelectronic devices, such as polarization-sensitive photodetectors, optical components, and polarized light-emitting diodes, and the linear dichroism can also be adopted to develop nanometric waveplates.

**RESULTS AND DISCUSSION**

**Crystal structure and sample preparation**

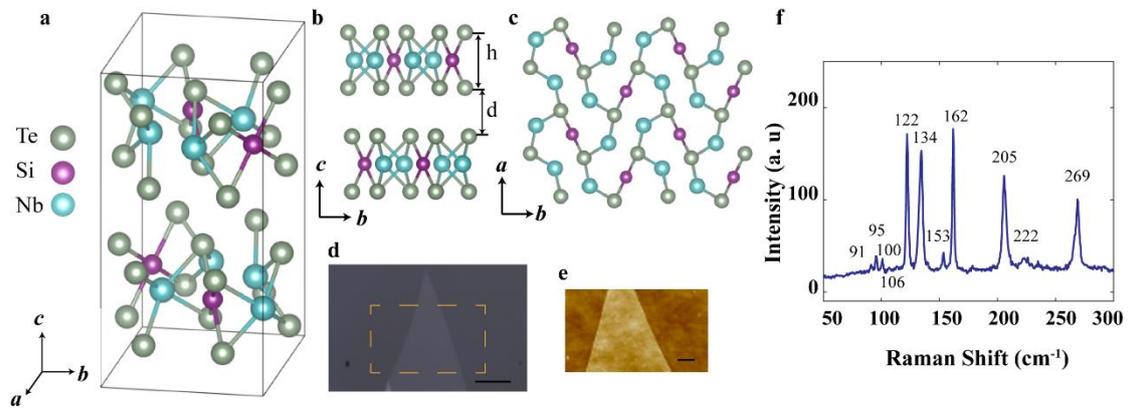

**Figure 1. Sample characterization. a** Illustration of few-layer $Nb_2SiTe_4$, showing atoms of Te (green), Nb (blue), and Si (purple). **b** Side view of few-layer $Nb_2SiTe_4$, $h = 3.783$ Å is the thickness of a sandwich layer, and $d = 3.310$ Å is the interlayer distance. **c** Top view of few-layer $Nb_2SiTe_4$. **d** Optical image of a $5\,nm$ $Nb_2SiTe_4$ on the quartz substrate. Scale bar, $10\,\mu m$. **e** AFM image of a $Nb_2SiTe_4$ sample marked by the orange dashed rectangle in (d), the thickness was determined to be $5\,nm$. Scale bar, $2\,\mu m$. **f** Raman spectrum of $Nb_2SiTe_4$.

The crystal structure of the NST is characterized by a stacking sequence as shown in Figure 1a. In this structure, with the symmetry of the space group $P2_1/c$, Nb atoms and Si atoms are sandwiched between two Te layers, and the Nb atoms and Si atoms are actually on the same plane. Besides, the two Te-Nb(Si)-Te sandwich layers per unit cell acquire an inversion symmetry with respect to the cell center. As reported in the literature, the thickness of a sandwich layer $h = 3.783$ Å, and the interlayer distance $d = 3.310$ Å, as indicated in Figure 1b. The sandwich layers are coupled by weak van der Waals interaction along *c*-axis, giving rise to the quasi-2D character of the electronic structure, and it allows bulk NST to be cleaved down to thin films by mechanical exfoliation. Figure 1d shows an optical image of a 5 nm thick flake on the quartz substrate exfoliated from a bulk crystal, which was synthesized by the self-flux method.[17] The thickness was determined by atomic force microscopy (AFM), as shown in Figure 1e. Figure 1f shows a representative Raman spectrum of NST with 532 nm laser excitation, in the range of 50-300 cm$^{-1}$. More details of Raman measurement of NST and the stability test based on it can be found in the Supporting Information.

**Anisotropic infrared response and corresponding inter-band transitions**

   **Infrared extinction spectra**

With a majority of the optical transitions expected in the infrared frequency range, Fourier transform infrared spectroscopy (FTIR) serves as the superior characterization technique. Infrared extinction spectra $(1 - \frac{T}{T_0})$ can be obtained to determine the band-band transitions, where $T$ is the light transmission from the sample and $T_0$ from the bare substrate, respectively (see Methods for more details). The typical polarization-resolved infrared extinction spectra of an 18 nm-thick sample at 10K are shown in Figure 2a, in the photon energy range from 0.46 to 1.5 eV. Beyond the lower boundary of 0.46 eV, there is no meaningful electronic transition feature. In low-dimensional materials, it is well known that the optical resonance peaks arise from excitons due to significantly reduced screening of Coulomb interaction. Especially when the temperature is low, the exciton resonances become even more salient. As shown in

Figure 2a, a sharp exciton peak appears. Variable temperature infrared measurements between 10K and 300K show monotonic decrease of the transition energies due to the electron-phonon coupling and lattice thermal expansion, which is common in semiconductors (see Supporting Information Figure S2). Infrared absorption study on NST indicates that pristine NST exhibits polarization-dependent resonance peaks, which is our major focus here. We assign the two features for polarization along *b*-axis to peaks A and B, and two features along *a*-axis to peaks B* and C, as indicated in Figure 2a. When the polarization changes from *b*-axis to *a*-axis, the lowest resonance peak A disappears, and a new peak C appears. Polarization-resolved extinction spectra with various light polarizations are shown in Figure S1. In addition, extinction spectra along *a*-axis and *b*-axis show a clear intersection at 0.66 eV. Among other existing 2D anisotropic materials which have also been demonstrated with linear dichroism, such as BP, ReX$_2$, AX (X=S and Se, and A=Ge, Sn)[27] and perovskite chalcogenides,[28] it is commonly observed that the absorption of light polarized along one crystal direction is always stronger than that along another crystal direction over a broad wavelength range. However, the linear dichroism polarity in NST shows a strong dependence on the wavelength. As shown in Figure 2a, *b*-axis extinction has much stronger intensity than that of *a*-axis in the photon energy range from 0.49 to 0.66 eV, and it behaves the opposite as the photon energy increases above 0.66 eV. Within the energy ranges below 0.49 eV and above 1.38 eV, *b*-axis spectrum tends to coincide with that of *a*-axis and the linear dichroism is much weaker. The intriguing linear dichroism switch phenomenon is also reported in the quasi-1D BaTiS$_3$,[28] which is readily used for polarization-wavelength selective photodetectors.

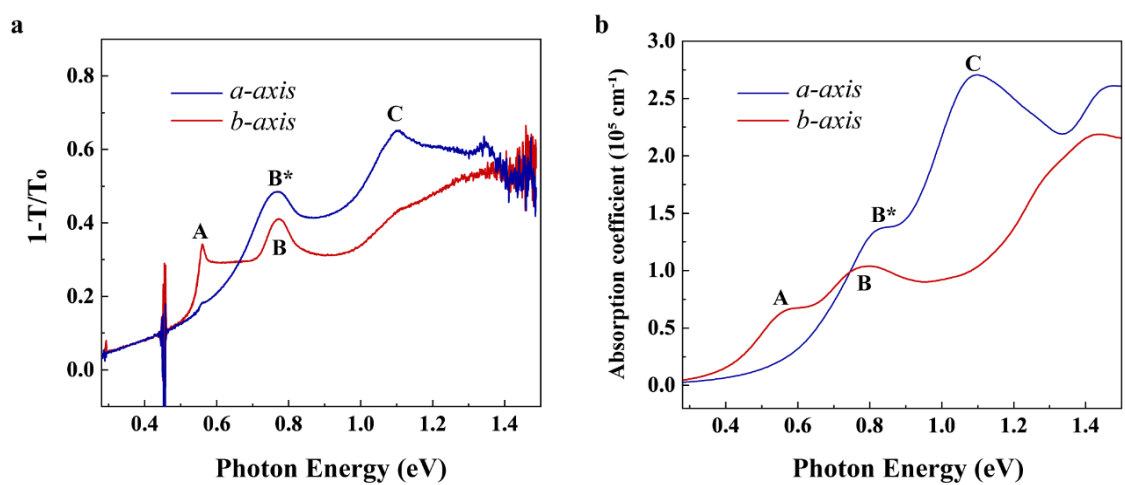

**Figure 2. Polarization-resolved extinction spectra in atomically thin NST. a** Infrared extinction spectra of an 18 nm NST for polarization along two different crystalline axes at 10K. **b** Polarization-resolved absorption coefficient using $G_0W_0$-BSE *ab initio* calculations.

**The origin of the anisotropic inter-band transitions**

To explore the origin of the anisotropic response and linear dichroism switch, first-principles calculations were performed. First, the PBE[29] and HSE06[30] functionals were applied to calculate the absorption spectra, but both failed to reproduce the optical resonance peaks that are observed experimentally (see Supporting Information Figure S8). Thus, we performed the one-shot Green's function ($G_0W_0$) and Bethe-Salpeter equation (BSE) calculation[31-33] for the bulk NST, which takes into account the excited states, such as the bound states of electron-hole pairs (excitons). As shown in Figure 2b, two prominent peaks are observed along the *a*-axis at 0.83 eV and 1.09 eV in the calculated spectra, which are consistent with the measured B* and C peaks in Figure

2a. On the other hand, along the *b*-axis direction, the calculated absorption spectrum shows peaks at 0.59 eV and 0.79 eV. We can see that the $G_0W_0$-BSE absorption spectra show very good agreement with the experimental infrared extinction spectra, which highlights the necessity to take into account the excitonic effect in the calculation of the optical response.

To further understand the origin of the peaks in the spectra, the electronic structure of NST was calculated using the HSE06 functional. Figure 3a and Figure 3b show the calculated band structures of bulk and monolayer NST, respectively, with band dispersion along the path $X - \Gamma - Y - R - X$ in the Brillouin zone. We can find that both bulk and monolayer NST have indirect bandgaps. The valance band maxima (VBMs) are located at the $\Gamma$ point, while the conduction band minima (CBMs) are located between $\Gamma$ and Y. Note that, the band dispersion along $\Gamma - Y$ and $\Gamma - X$ directions are different. Along the $\Gamma - Y$ direction, it shows relatively small dispersion compared to that along $\Gamma - X$, which certainly implies anisotropic infrared absorption features. The calculated bandgap of bulk NST is about 0.39 eV, and the bandgap of monolayer NST is about 0.89 eV, in agreement with previous experimental measurements[17] and theoretical predictions,[16] which validates our calculation results. The band structures of few-layer NST can be found in the Supporting Information Figure S5 and no significant change occurs, except that the bandgap increases as the number of layers decreases due to the quantum confinement effect.[34] The corresponding spectral shape in the experiment does not change much either in the thicker samples (see Supporting Information Figure S6).

Now, let us revisit the anisotropic transition features. In particular, we focus on the polarization dependence of resonance peaks, which are mainly originated from direct transitions at Γ-point of the Brillouin zone, where the joint densities of states (JDOS) are large. To understand the optical selection rules, we performed group theory analysis. From the perspective of group theory, the optical selection rules are interpreted as telling whether the matrix element $\left(\psi_\alpha^{(i)}, \mathcal{H}'\psi_{\alpha'}^{(i')}\right)$ vanishes or not through symmetry analysis,[35] where $\mathcal{H}'$ is the interaction Hamiltonian matrix that couples the final states $\psi_\alpha$ and initial state $\psi_{\alpha'}$. The superscript $i(i')$ is the abbreviated notation for the irreducible representation $\Gamma_{i(i')}$. The Hamiltonian $\mathcal{H}'$ can be expressed in terms of the irreducible representations in the group of Schrödinger's equation:

$$\mathcal{H}' = \sum_{j,\beta} f_\beta^{(j)} \mathcal{H}'_\beta^{(j)} \tag{1}$$

where $j$ denotes the irreducible representations $\Gamma_j$ of the Hamiltonian $\mathcal{H}'$, and $\beta$ denotes the partners of $\Gamma_j$. The transitions are prohibited by symmetry if and only if the direct product $\psi_\alpha^{(i)} \otimes \mathcal{H}'_\beta^{(j)} \otimes \psi_{\alpha'}^{(i')}$ does not contain the identity irreducible representation. Equivalently, the transitions are prohibited if and only if $\psi_\alpha^{(i)}$ is orthogonal to all the basis functions in the decomposition of $\mathcal{H}'\psi_{\alpha'}^{(i')}$ into irreducible representations.

**Table 1.** Character table of the Γ-point space group. $x$, $y$, and $z$ represent the components of polar vectors. $c_1$ to $c_5$ and $v_1$ to $v_3$ represent the conduction bands and the valence bands in the Γ-point area, respectively.

|  | $E$ | $\tau C_{2y}$ | $i$ | $\tau m_y$ |  |  |
| --- | --- | --- | --- | --- | --- | --- |
| $\Gamma_1^+$ | 1 | 1 | 1 | 1 | | $c_1, c_3, v_2$ |
| $\Gamma_2^+$ | 1 | -1 | 1 | -1 | | $c_5$ |
| $\Gamma_1^-$ | 1 | 1 | -1 | -1 | $y$ | $c_2, c_4, v_1$, |
| $\Gamma_2^-$ | 1 | -1 | -1 | 1 | $x, z$ | $v_3$ |

NST has the symmetry with the space group of P2$_1$/c. The point group at Γ-point is $C_{2h}$. The four symmetry operations and four irreducible representations are listed in Table 1. Furthermore, we also determined the irreducible representations of three valence bands (denoted as $v_1$, $v_2$, $v_3$ in Figure 3d) and five conduction bands (denoted as $c_1$, $c_2$, $c_3$, $c_4$, $c_5$ in Figure 3d) near the band edge from their calculated wavefunctions. The electromagnetic interaction Hamiltonian giving rise to the electric dipole transitions reads: [35]

$$\mathcal{H}' = -\frac{e}{mc} \boldsymbol{p} \cdot \boldsymbol{A} \qquad (2)$$

where $\boldsymbol{p}$ is the momentum of the electron and $\boldsymbol{A}$ is the vector potential of the external electromagnetic field. Under the symmetry operations, $\boldsymbol{A}$ is invariant. $p_x$ and $p_z$ transform as the irreducible representation $\Gamma_2^-$, and $p_y$ transforms as the irreducible representation $\Gamma_1^-$ as indicated in Table 1. If the initial state is $v_1$ with $\Gamma_1^-$ and the polarization direction is $x$ ($a$-axis) with $\Gamma_2^-$, we can take the direct product $\Gamma_1^- \otimes \Gamma_2^- = \Gamma_2^+$. Thus, the selection rule requires that the final state can only be the states with $\Gamma_2^+$, i.e. $c_5$ state. Similarly, one can use the above arguments to work out the selection rules for all the initial valence states and for both $x$ ($a$-axis) and $y$ ($b$-axis) polarizations. The results are shown in Figure 3c. More details on group theory analysis can be found in Supporting Information Note 3.

Furthermore, based on the group theory analysis and the calculated inter-band optical transition matrix elements (MEs), we assign the peaks along *a*-axis and *b*-axis to the optical transitions as marked in Figure 2. Along the *a*-axis, peaks B* and C are assigned to the transitions $v_1 \rightarrow c_3$ and $v_1 \rightarrow c_5$, respectively. Along the *b*-axis, peaks A and B are assigned to $v_1 \rightarrow c_1$ and $v_1 \rightarrow c_3$, respectively, as shown in Figure 3d. One should note that the transition between bands $v_1 \rightarrow c_3$ (corresponding to the peak B*) is not allowed at Γ-point. The calculated MEs suggest that peak B* should be assigned to the transition between $v_1 \rightarrow c_3$ at a point between Γ − X rather than the exact Γ-point.

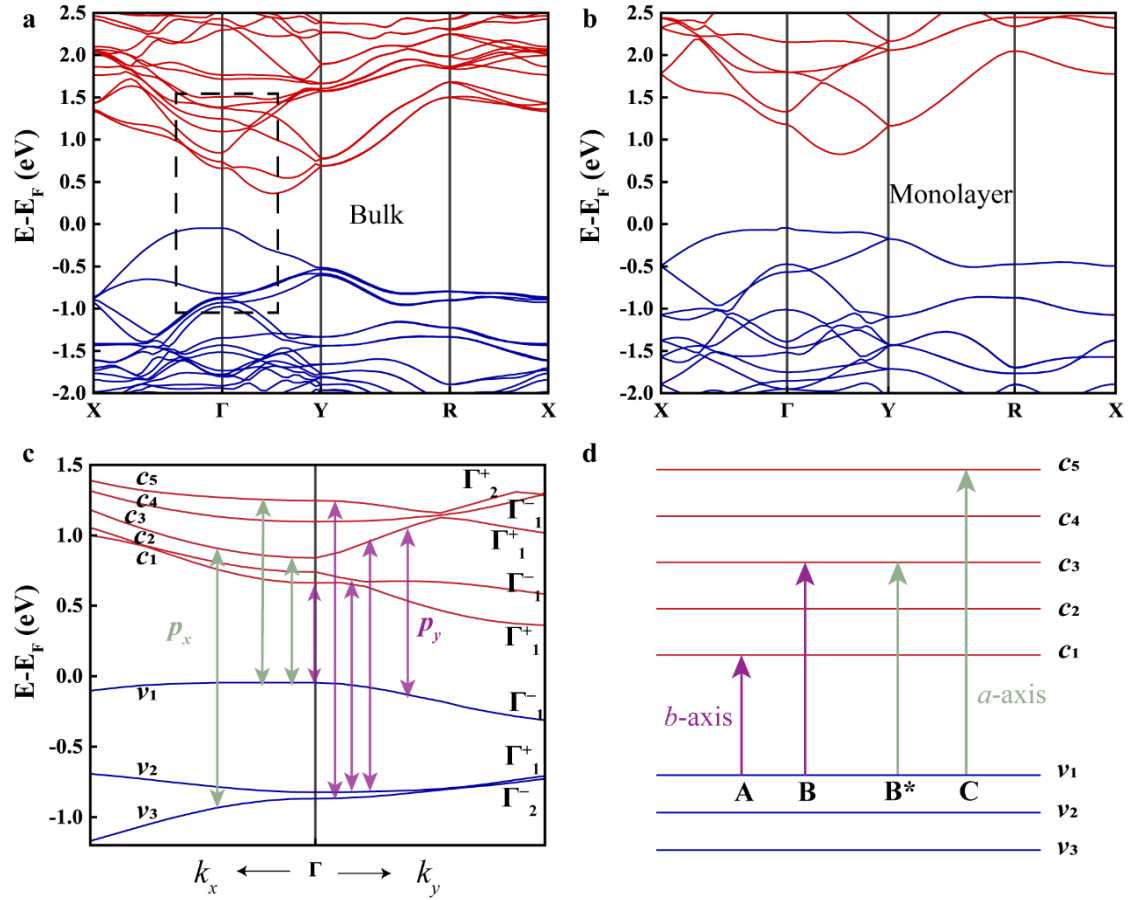

**Figure 3. Band structures and optical transition selection rules.** Calculated band structures of **a** Bulk NST and **b** monolayer NST. **c** The band structure inside the black dashed rectangle of **a**, illustrating selection rules allowed by symmetry. Only electronic bands involved in the transitions are shown, namely, the bands of $c_1$ to $c_5$ and $v_1$ to $v_3$. Light green double arrows indicate the transitions which are allowed along the *a*-axis polarization. The purple double arrows indicate the permitted transitions along *b*-axis polarization. **d** Schematic illustrations of optical transitions. B* and C denote transitions $v_1 \rightarrow c_3$ and $v_1 \rightarrow c_5$ along *a*-axis polarization, respectively. A and B denote transitions $v_1 \rightarrow c_1$ and $v_1 \rightarrow c_3$ along *b*-axis polarization, respectively. The rest of transitions in **c** are out of the energy range of our measurements.

**Band-structure engineering.**

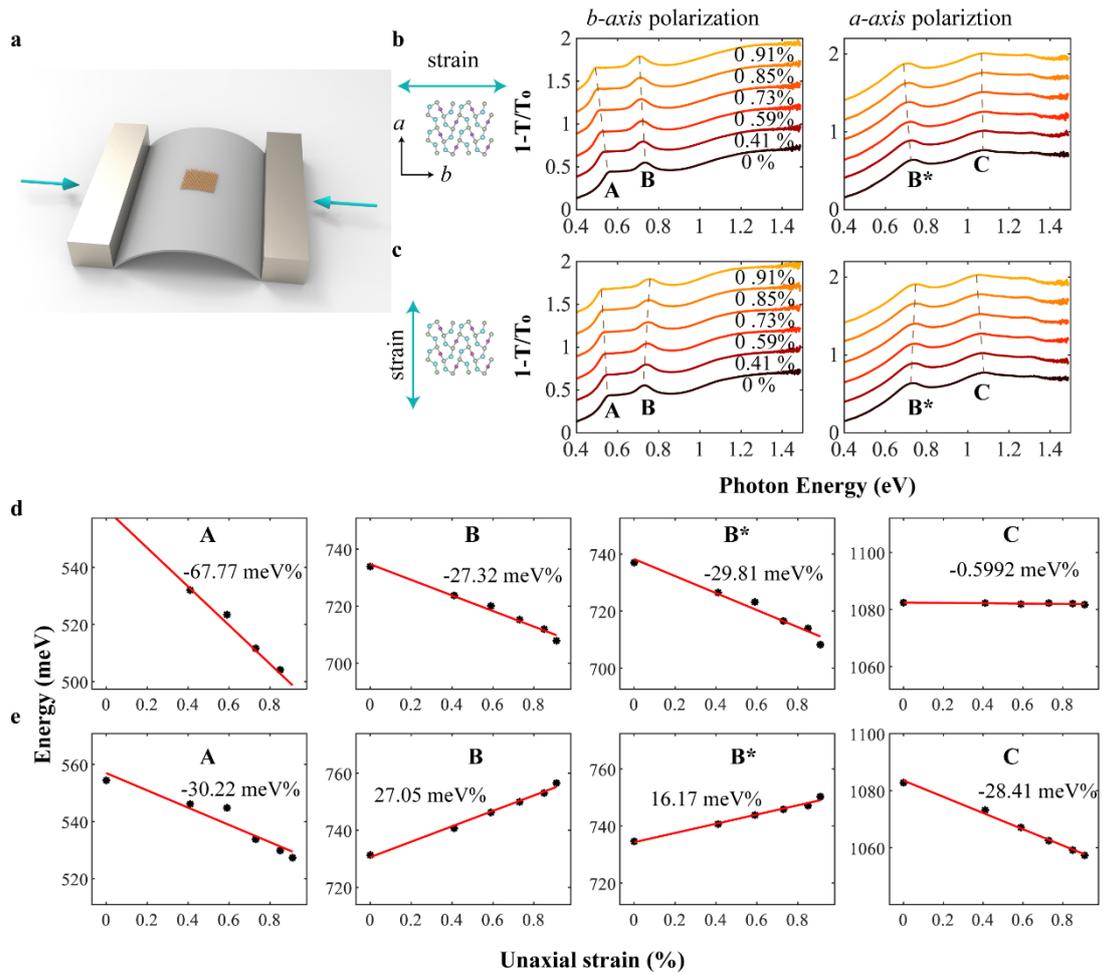

**Figure 4. Strain engineering. a** Schematic illustration of the two-point bending apparatus. The infrared extinction spectra under varying tensile strains along **b** *b*-axis tensile strain and **c** *a*-axis tensile strain. For clarity, the spectra are vertically offset. Dashed lines are guides to the eye. The left panel is the extinction spectra under *b*-axis polarization, and the right one is under *a*-axis polarization. Transition energies as a function of tensile strain for **d** strain along *b*-axis and **e** strain along *a*-axis.

After discussing the inter-band optical transitions, we now focus on the tuning of the band structure, which shows even more intriguing scenarios. The electronic structure of NST has been predicted to be very sensitive to strain[22] and the optical transitions of NST are polarization-dependent. We thus expect the strain effects are crystallographic orientation-dependent. By employing a two-point bending apparatus to induce controllable uniaxial strain,[36] as illustrated in Figure 4a (more details on the strain value quantification can be found in the Supporting Information Note 1), we traced the band-structure evolution. Before applying strain, the crystallographic orientation of the NST flake was identified by polarized infrared spectroscopy, so that the uniaxial strain could be applied in the desired direction. As shown in Figure 4b and 4c, the characteristic peaks show different dependence on strain applied along two different axes. The peak positions in Figure 4b and 4c are summarized in Figure 4d and 4e as a function of tensile strain along both *b*-axis and *a*-axis. With *b*-axis strain, all the transition resonance peaks redshift, though with different shift rates. The redshift under tensile strain is typically observed in other semiconductors, such as $MoS_2$.[37] It can be qualitatively understood as a result of reduced orbital overlap and hybridization due to weakened atomic bonds. Interestingly, opposite changes are observed for both peaks B and B* with strain along *a*-axis (additional data can be found in the Supporting Information Figure S7). The unusual blueshift of resonance energies is also observed in few-layer BP,[38] which is the result of the competition between two intralayer hopping parameters, one of which is negative. It may be natural that the anisotropic 2D material exhibits the crystallographic orientation-dependent strain effect.[39] However, the opposite shift along two strain directions cannot be taken for granted. For instance, few-layer BP even shows almost no dependence on the strain direction.[3] Given the opposite shift with strain along two principal axes, in principle, there exists a special direction, along which the strain shows no effect on B peak or B* peak. This may find application in strain-immune devices.

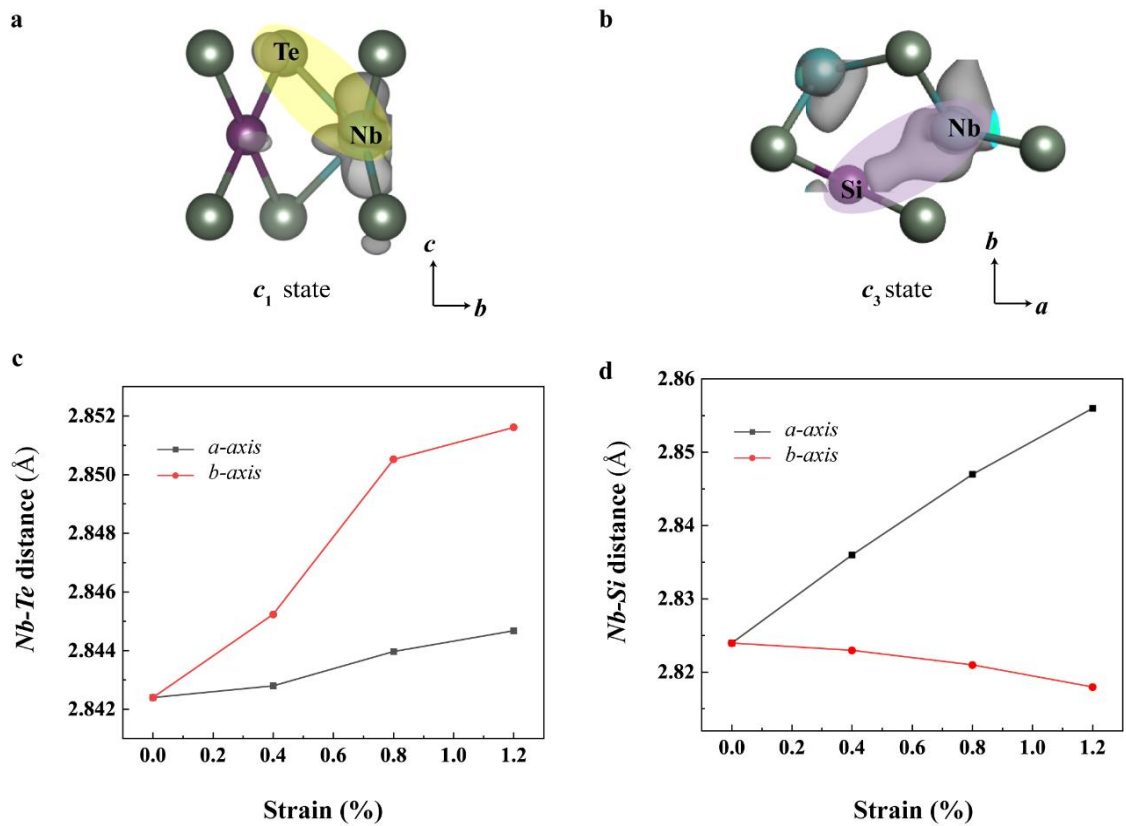

**Figure 5. Orbital coupling analysis. a** Calculated partial charge distribution of the $c_1$ state from its wavefunction (Nb-Te dominated). **b** Calculated partial charge distribution of the $c_3$ state (Nb-Si dominated). The gray surfaces are the iso-surface levels of charge densities. **c** The Nb-Te bond length versus uniaxial tensile strain along two directions. **d** the Nb-Si bond length as a function of uniaxial strain along two directions.

At this point, it is very revealing to qualitatively interpret some of the aforementioned experimental results by the computed lattice constants and bond characteristics. To understand the strain effects of the characteristic peaks, we performed calculations of the structure relaxations with different uniaxial strains, and calculated the partial charge densities of the electronic bands and further analyzed the orbital compositions (see Methods). The composition of $v_1$ state, which serves as the initial state for the transitions of all the four characteristic peaks, is dominated by the Nb 4$d$ orbitals (84% contribution), suggesting the orbital coupling is weak and only slightly affected by strain. However, the conduction bands involved in the transitions (e.g. $c_1$, $c_3$, $c_5$) show strong orbital couplings, the strengths of which are affected by uniaxial strain. For peak A, the final state $c_1$ is characterized by the coupling of Nb $d_{z^2}$ and Te $p_x$, $p_y$ orbitals. The partial charge density is shown in Figure 5a. We can notice that the charge resides outside of the bond, indicating an antibonding state of Nb-Te hybridization orbitals. Figure 5c shows the calculated Nb-Te bond length under various strain along two principal directions. We can see that the length of Nb-Te bond all increases. Thus, the coupling strength is weakened, and the $c_1$ state will show the tendency of down-shift. Given that the charge distribution of $v_1$ state is mostly located at isolated Nb atoms, the $v_1$ state shows less dependence on the uniaxial strain. As a result, the peak A exhibits redshift when the uniaxial tensile strain is applied (the same applies to peak C, see Supporting Information Note 4).

However, for both peaks B and B*, opposite shifts were observed. To explain this unusual phenomenon, let us take a closer look at the partial charge density of $c_3$, the final state of peaks B and B*. The $c_3$ state is hybridized by the Nb $d_{x^2-y^2}$ and Si $s$ orbitals, as shown in Figure 5b. We also notice that most of the charges are located between the Nb and Si atoms, forming a covalent bond, so the $c_3$ state acts as a bonding state of Nb-Si hybridization orbitals. The trend of Nb-Si bond length as a function of strain is plotted in Figure 5d. When a 1.2% tensile strain is applied along $a$-axis, the

Nb-Si bond length increases from 2.824 Å to 2.856 Å. Thus, the strength of the Nb-Si bond and the *p-d* coupling become weaker, and the energy of bonding state $c_3$ is pushed higher, giving rise to blueshifts of peaks B and B*. However, for the tensile strain along *b*-axis, the Nb-Si bond length decreases (Figure 5d), which enhances the Nb-Si bond, resulting in redshifts of both peaks.

**CONCLUSIONS**

In summary, we have comprehensively investigated the optical inter-band transitions in NST. The in-plane anisotropic optical transitions were observed by infrared extinction spectra and well explained by the DFT calculation and group theory analysis. Most importantly, the observed orientation-dependent strain-induced peak shifts are consistent with analysis based on the hybridized orbitals. We believe that our comprehensive study will stimulate further research efforts on tunable anisotropic optoelectronic devices.

**METHODS**

*Sample preparation.* The bulk NST crystal was synthesized with self-flux method. Thin NST film samples were prepared by a polydimethylsiloxane (PDMS) assisted mechanical exfoliation method. First, the NST flakes were cleaved on a PDMS substrate, then transferred onto a quartz substrate for further optical measurements once thin flakes were identified under a microscope.

*Polarization-resolved Infrared spectroscopy.* The polarization-resolved infrared measurements on thin NST films were performed using a Bruker FTIR spectrometer (Vertex 70v) equipped with a Hyperion 2000 microscope. A tungsten halogen lamp was used as the light source, in combination with a liquid nitrogen-cooled mercury-cadmium-telluride (MCT) detector. The incident light was focused on NST flakes using a 15X infrared condenser, and the transmitted light was collected by another condenser with the same specifications. The polarization was controlled by a broadband ZnSe grid polarizer.

*Raman spectroscopy.* Raman measurements were conducted using a commercial Raman system (Horiba, Jobin Yvon HR-Evolution 2) with 532 nm laser for excitation, using 100X objective lens (NA=0.9) and 1800 grooves per mm grating. All the spectra were calibrated in energy using the silicon peak at 520.7 cm$^{-1}$.

*DFT Calculations.* All the theoretical calculations including structure relaxation, band structure calculation, and $G_0W_0$-BSE calculation were performed via the Vienna *ab initio* simulation package (VASP).[40] The projected augmented-wave (PAW) pseudopotentials were used to treat the core and valence electrons,[41] with 11 electrons for Nb, 4 electrons for Si and 6 electrons for Te. The Heyd-Scuseria-Ernzerhof (HSE) hybrid functional was applied,[30] with 25% of the Perdew-Burke-Ernzerhof (PBE) exchange functional replaced by the screened Hartree-Fock exchange. A 28-atom unit cell for the bulk NST and a 14-atom unit cell with a 15 Å vacuum slab for monolayer NST were used in the calculations. The cutoff energy of the plane-wave basis is 400 eV. The Brillion zone was sampled with a $5 \times 4 \times 2$ Γ-centered *k*-point mesh and the electronic minimization was performed with a tolerance of $10^{-6}$ eV. The van der Waals interactions between layers were considered using zero damping DFT-D3 method of Grimme as implemented in VASP.[42-43] For the structure relaxation, the atomic structures were relaxed until the atomic forces are smaller than 0.005 eV·Å$^{-1}$, and the *c* axis of the lattice parameter (out of plane direction) was fixed for the monolayer and multilayer structure relaxation.

For the G$_0$W$_0$-BSE calculation,[31-33] we set a cutoff of 250 eV for the response function and the number of frequency grid points were set to 240. The G$_0$W$_0$-BSE calculation was conducted as follows: firstly, we obtained the electronic structure of the NST ground state through common DFT calculation; secondly, we obtained the correction of the quasiparticle effect through solving the self-energy in a single shot, which was approximated by the product of the one-particle Green's function *G* and the screened Coulomb potential *W* (the many-body exchange-correlation interactions were included) ; thirdly, based on the quasiparticle corrected energy and wavefunctions, we performed the BSE calculation, which described the interactions of electron-hole pairs and obtained the optical excitation energies by self-consistently solving the BSE:[44-45]

$$(E_{c\boldsymbol{k}} - E_{v\boldsymbol{k}})A_{vc\boldsymbol{k}}^S + \sum_{\boldsymbol{k}'v'c'} \langle vc\boldsymbol{k}|K_{eh}|v'c'\boldsymbol{k}'\rangle A_{v'c'\boldsymbol{k}'}^S = \Omega^S A_{vc\boldsymbol{k}}^S. \qquad (3)$$

Here, $\Omega^S$ is the exciton energy. $E_{c\boldsymbol{k}}$ ($E_{v\boldsymbol{k}}$) is the energy of the conduction (valence) band from the second step. $A_{vc\boldsymbol{k}}^S$ is the expansion coefficients for the exciton wave function $|vc\boldsymbol{k}\rangle$, and $K_{eh}$ is the electron-hole interaction kernel. The dielectric functions were obtained following reference,[46] except that GW quasiparticle energies instead of DFT energies were used. Then the absorption spectrum α(ω) was

calculated by[47]

$$\alpha(\omega) = \frac{\sqrt{2}\omega}{c}\left[\sqrt{\varepsilon_r^2 + \varepsilon_i^2} - \varepsilon_r\right]^{\frac{1}{2}} \quad (4)$$

where $\varepsilon_r$ and $\varepsilon_i$ are the real and imaginary parts of the dielectric function along the high symmetry directions (*a*- or *b*-axis), and $\omega$ is the photon energy.

The projected wave function character of each orbital was evaluated by calculating the projections for every ion $|\langle Y_{lm}^{\alpha}|\phi_{nk}\rangle|^2$, where $Y_{lm}^{\alpha}$ is the spherical harmonic centered at ion index $\alpha$, *l* and *m* are the angular moment and magnetic quantum. $\phi_{nk}$ is the wavefunction. *n* and *k* are the band and k-point index, respectively.

*Strain set-up and strain-dependent Infrared measurements.* To induce controllable uniaxial strain, we used flexible PP substrate and employed a two-point bending apparatus. First, the PP substrate was cut into a square shape with $4 \text{ cm} \times 4 \text{ cm}$. Second, the crystallographic orientation of the NST flake was identified by polarized infrared spectroscopy, so that the uniaxial strain could be applied to the sample in the desired directions (*a*-axis or *b*-axis). Finally, the sample was transferred onto the central of the square PP substrate with fixed direction. The applied strain was kept below 1% and the process was reversible and repeatable (see Supporting Information Note 1). We traced the Infrared spectra by bending the PP substrate along *a*-axis (*b*-axis), then gradually released the strain.

**ASSOCIATED CONTENT**

**Supporting Information**

The Supporting Information is available at *** (PDF)

Details of theoretical calculations, strain setup, various strain-tunable samples, the stability test of NST (time-evolution of the infrared extinction spectra and Raman spectra), laser power- and polarization-dependent Raman spectra.

**Author Contributions**

Fanjie Wang and Yonggang Xu contributed equally to this work.

**Notes**

The authors declare no competing financial interest.

**ACKNOWLEDGEMENTS**

H.G.Y. is grateful to the financial support from the National Natural Science Foundation of China (Grant Nos. 11874009, 11734007), the National Key Research and Development Program of China (Grant Nos. 2021YFA1400100, 2017YFA0303504), the Natural Science Foundation of Shanghai (Grant No. 20JC1414601), and the Strategic Priority Research Program of Chinese Academy of Sciences (XDB30000000). J. -H. Y. is grateful to the National Natural Science Foundation of China (Grants Nos. 11991061, 11974078), and Shanghai Sailing Program (Grant No. 19YF1403100). Computations were performed at the High-Performance Computing Center of Fudan University. Y.F.G. acknowledges research funds from the State Key Laboratory of Surface Physics and Department of Physics, Fudan University (Grant No. KF2020_09).

# REFERENCES

1. Li, L.; Kim, J.; Jin, C.; Ye, G. J.; Qiu, D. Y.; da Jornada, F. H.; Shi, Z.; Chen, L.; Zhang, Z.; Yang, F.; Watanabe, K.; Taniguchi, T.; Ren, W.; Louie, S. G.; Chen, X. H.; Zhang, Y.; Wang, F., Direct Observation of the Layer-Dependent Electronic Structure in Phosphorene. *Nat. Nanotechnol.* **2016,** *12* (1), 21-25.
2. Tran, V.; Soklaski, R.; Liang, Y. F.; Yang, L., Layer-Controlled Band Gap and Anisotropic Excitons in Few-Layer Black Phosphorus. *Phys. Rev. B* **2014,** *89* (23), 235319.
3. Zhang, G.; Huang, S.; Chaves, A.; Song, C.; Ozcelik, V. O.; Low, T.; Yan, H., Infrared Fingerprints of Few-Layer Black Phosphorus. *Nat. Commun.* **2017,** *8*, 14071.
4. Li, P.; Appelbaum, I., Electrons and Holes in Phosphorene. *Phys. Rev. B* **2014,** *90* (11), 115439.
5. Zhang, G.; Huang, S.; Wang, F.; Xing, Q.; Song, C.; Wang, C.; Lei, Y.; Huang, M.; Yan, H., The Optical Conductivity of Few-Layer Black Phosphorus by Infrared Spectroscopy. *Nat. Commun.* **2020,** *11* (1), 1847.
6. Zhang, G.; Huang, S.; Wang, F.; Xing, Q.; Low, T.; Yan, H., Determination of Layer-Dependent Exciton Binding Energies in Few-Layer Black Phosphorus. *Sci. Adv.* **2018,** *4*, eaap9977.
7. Qiao, J.; Kong, X.; Hu, Z. X.; Yang, F.; Ji, W., High-Mobility Transport Anisotropy and Linear Dichroism in Few-Layer Black Phosphorus. *Nat. Commun.* **2014,** *5*, 4475.
8. Nemilentsau, A.; Low, T.; Hanson, G., Anisotropic 2D Materials for Tunable Hyperbolic Plasmonics. *Phys. Rev. Lett.* **2016,** *116* (6), 066804.
9. Edo, V. V.; Andrei, N.; Anshuman, K.; Rafael, R.; Mikhail I. Katsnelson; Tony, L.; Shengjun, Y., Tuning Two-Dimensional Hyperbolic Plasmons in Black Phosphorus. *Phys. Rev. A* **2019,** *12* (1), 014011.
10. Correas-Serrano, D.; Gomez-Diaz, J. S.; Melcon, A. A.; Alù, A., Black Phosphorus Plasmonics: Anisotropic Elliptical Propagation and Nonlocality-Induced Canalization. *J. Opt.* **2016,** *18* (10), 104006.
11. Wang, F.; Wang, C.; Chaves, A.; Song, C.; Zhang, G.; Huang, S.; Lei, Y.; Xing, Q.; Mu, L.; Xie, Y.; Yan, H., Prediction of Hyperbolic Exciton-Polaritons in Monolayer Black Phosphorus. *Nat. Commun.* **2021,** *12*, 5628.
12. Favron, A.; Gaufres, E.; Fossard, F.; Phaneuf-L'Heureux, A. L.; Tang, N. Y.; Levesque, P. L.; Loiseau, A.; Leonelli, R.; Francoeur, S.; Martel, R., Photooxidation and Quantum Confinement Effects in Exfoliated Black Phosphorus. *Nat. Mater.* **2015,** *14* (8), 826-832.
13. Zhou, Q.; Chen, Q.; Tong, Y.; Wang, J., Light-Induced Ambient Degradation of Few-Layer Black Phosphorus: Mechanism and Protection. *Angew. Chem. Int. Ed.* **2016,** *55* (38), 11437-11441.
14. Wang, F.; Zhang, G.; Huang, S.; Song, C.; Wang, C.; Xing, Q.; Lei, Y.; Yan, H., Electronic Structures of Air-Exposed Few-Layer Black Phosphorus by Optical Spectroscopy. *Phys. Rev. B* **2019,** *99* (7), 075427.
15. Mak, K. F.; Lee, C.; Hone, J.; Shan, J.; Heinz, T. F., Atomically Thin $MoS_2$: A New Direct-Gap Semiconductor. *Phys. Rev. Lett.* **2010,** *105* (13), 136805.
16. Tremel, W.; Kleinke, H.; Derstroff, V.; Reisner, C., Transition Metal Chalcogenides: New Views on an Old Topic. *J. Alloys Compd.* **1995,** *219* (1-2), 73-82.
17. Zhao, M.; Xia, W.; Wang, Y.; Luo, M.; Tian, Z.; Guo, Y.; Hu, W.; Xue, J., $Nb_2SiTe_4$: A Stable Narrow-Gap Two-Dimensional Material with Ambipolar Transport and Mid-Infrared Response.

*ACS Nano* **2019,** *13* (9), 10705-10710.

18. Fang, W.-Y.; Li, P.-A.; Yuan, J.-H.; Xue, K.-H.; Wang, J.-F., Nb$_2$SiTe$_4$ and Nb$_2$GeTe$_4$: Unexplored 2d Ternary Layered Tellurides with High Stability, Narrow Band Gap and High Electron Mobility. *J. Electro.Mater.* **2019,** *49* (2), 959-968.

19. Sajjad, M.; Singh, N., The Impact of Electron–Phonon Coupling on the Figure of Merit of Nb$_2$SiTe$_4$ and Nb$_2$GeTe$_4$ Ternary Monolayers. *Phys. Chem. Chem. Phys.* **2021,** *23* (29), 15613-15619.

20. Wang, B.; Xia, W.; Li, S.; Wang, K.; Yang, S. A.; Guo, Y.; Xue, J., One-Dimensional Metal Embedded in Two-Dimensional Semiconductor in Nb$_2$Si$_{k-1}$te$_4$. *ACS Nano* **2021,** *15*, 7149-7154.

21. Zhou, K.; Deng, J.; Chen, L.; Xia, W.; Guo, Y.; Yang, Y.; Guo, J.-G.; Guo, L., Observation of Large in-Plane Anisotropic Transport in Van Der Waals Semiconductor Nb2site4. *Chin. Phys. B* **2021,** *30*, 087202.

22. Zhang, T.; Ma, Y.; Xu, X.; Lei, C.; Huang, B.; Dai, Y., Two-Dimensional Ferroelastic Semiconductors in Nb2site4 and Nb2gete4 with Promising Electronic Properties. *J. Phys. Chem. Lett.* **2020,** *11* (2), 497-503.

23. Wu, X.; Gao, G. Y.; Hu, L.; Qin, D., 2D Nb$_2$SiTe$_4$ and Nb$_2$SiTe$_4$: Promising Thermoelectric Figure of Merit and Gate-Tunable Thermoelectric Performance. *Nanotechnology* **2021,** *32*, 245203.

24. Hu, J.; Liu, X.; Yue, C. L.; Liu, J. Y.; Zhu, H. W.; He, J. B.; Wei, J.; Mao, Z. Q.; Antipina, L. Y.; Popov, Z. I.; Sorokin, P. B.; Liu, T. J.; Adams, P. W.; Radmanesh, S. M. A.; Spinu, L.; Ji, H.; Natelson, D., Enhanced Electron Coherence in Atomically Thin nb3site6. *Nat. Phys.* **2015,** *11* (6), 471-476.

25. Li, S.; Liu, Y.; Wang, S.-S.; Yu, Z.-M.; Guan, S.; Sheng, X.-L.; Yao, Y.; Yang, S. A., Nonsymmorphic-Symmetry-Protected Hourglass Dirac Loop, Nodal Line, and Dirac Point in Bulk and Monolayer X$_3$SiTe$_6$ (X = Ta, Nb). *Phys. Rev. B* **2018,** *97* (4), 045131.

26. Yao, J.; Yang, G., Multielement 2D Layered Material Photodetectors. *Nanotechnology* **2021,** *32*, 392001.

27. Su, J.; Shen, W.; Chen, J.; Yang, S.; Liu, J.; Feng, X.; Zhao, Y.; Hu, C.; Li, H.; Zhai, T., 2D Ternary Vanadium Phosphorous Chalcogenide with Strong in-Plane Optical Anisotropy. *Inorg. Chem. Front.* **2021,** *8*, 2999.

28. Wu, J.; Cong, X.; Niu, S.; Liu, F.; Zhao, H.; Du, Z.; Ravichandran, J.; Tan, P. H.; Wang, H., Linear Dichroism Conversion in Quasi-1D Perovskite Chalcogenide. *Adv. Mater.* **2019,** *31* (33), e1902118.

29. Perdew, J. P.; Burke, K.; Ernzerhof, M., Generalized Gradient Approximation Made Simple. *Phys. Rev. Lett.* **1996,** *77* (18), 3865-3868.

30. Heyd, J.; Scuseria, G. E.; Ernzerhof, M., Hybrid Functionals Based on a Screened Coulomb Potential. *J. Chem. Phys.* **2003,** *118* (18), 8207-8215.

31. Shishkin, M.; Kresse, G., Implementation and Performance of the Frequency-Dependent $Gw$ Method within the Paw Framework. *Phys. Rev. B* **2006,** *74* (3), 035101.

32. Shishkin, M.; Marsman, M.; Kresse, G., Accurate Quasiparticle Spectra from Self-Consistent Gw Calculations with Vertex Corrections. *Phys. Rev. Lett.* **2007,** *99* (24), 246403.

33. Shishkin, M.; Kresse, G., Self-Consistent $Gw$ Calculations for Semiconductors and Insulators. *Phys. Rev. B* **2007,** *75* (23), 235102.

34. Takagahara, T.; Takeda, K., Theory of the Quantum Confinement Effect on Excitons in Quantum Dots of Indirect-Gap Materials. *Phys. Rev. B* **1992,** *46* (23), 15578-15581.

35. Dresselhaus, M. S.; Dresselhaus, G.; Jorio, A., *Group Theory: Application to the Physics of Condensed Matter*. Springer Science & Business Media: Berlin, 2007, 97-107.

36. Song, C.; Fan, F.; Xuan, N.; Huang, S.; Wang, C.; Zhang, G.; Wang, F.; Xing, Q.; Lei, Y.; Sun, Z.;

Wu, H.; Yan, H., Drastic Enhancement of the Raman Intensity in Few-Layer Inse by Uniaxial Strain. *Phys. Rev. B* **2019,** *99* (19), 195414.

37. Conley, H. J.; Wang, B.; Ziegler, J. I.; Haglund, R. F., Jr.; Pantelides, S. T.; Bolotin, K. I., Bandgap Engineering of Strained Monolayer and Bilayer $MoS_2$. *Nano Lett.* **2013,** *13* (8), 3626-3630.

38. Huang, S.; Zhang, G.; Fan, F.; Song, C.; Wang, F.; Xing, Q.; Wang, C.; Wu, H.; Yan, H., Strain-Tunable Van Der Waals Interactions in Few-Layer Black Phosphorus. *Nat. Commun.* **2019,** *10* (1), 2447.

39. Li, H.; Sanchez-Santolino, G.; Puebla, S.; Frisenda, R.; Al-Enizi, A. M.; Nafady, A.; D'Agosta, R.; Castellanos-Gomez, A., Strongly Anisotropic Strain-Tunability of Excitons in Exfoliated Zrse3. *Adv. Mater.* **2021,** *34*, e2103571.

40. Kresse, G.; Furthmüller, J., Efficient Iterative Schemes for Ab Initio Total-Energy Calculations Using a Plane-Wave Basis Set. *Phys. Rev. B* **1996,** *54* (16), 11169-11186.

41. Blöchl, P. E., Projector Augmented-Wave Method. *Phys. Rev. B* **1994,** *50* (24), 17953-17979.

42. Grimme, S.; Antony, J.; Ehrlich, S.; Krieg, H., A Consistent and Accurate Ab Initio Parametrization of Density Functional Dispersion Correction (DFT-D) for the 94 Elements H-Pu. *J. Chem. Phys.* **2010,** *132* (15), 154104.

43. Grimme, S.; Ehrlich, S.; Goerigk, L., Effect of the Damping Function in Dispersion Corrected Density Functional Theory. *J. Comput. Chem.* **2011,** *32* (7), 1456-1465.

44. Onida, G.; Reining, L.; Rubio, A., Electronic Excitations: Density-Functional Versus Many-Body Green's-Function Approaches. *Rev. Mod. Phys.* **2002,** *74* (2), 601-659.

45. Rohlfing, M.; Louie, S. G., Electron-Hole Excitations and Optical Spectra from First Principles. *Phys. Rev. B* **2000,** *62* (8), 4927-4944.

46. Gajdoš, M.; Hummer, K.; Kresse, G.; Furthmüller, J.; Bechstedt, F., Linear Optical Properties in the Projector-Augmented Wave Methodology. *Phys. Rev. B* **2006,** *73* (4), 045112.

47. Huang, X.; Paudel, T. R.; Dong, S.; Tsymbal, E. Y., Hexagonal Rare-Earth Manganites as Promising Photovoltaics and Light Polarizers. *Phys. Rev. B* **2015,** *92* (12), 125201.

**TOC**

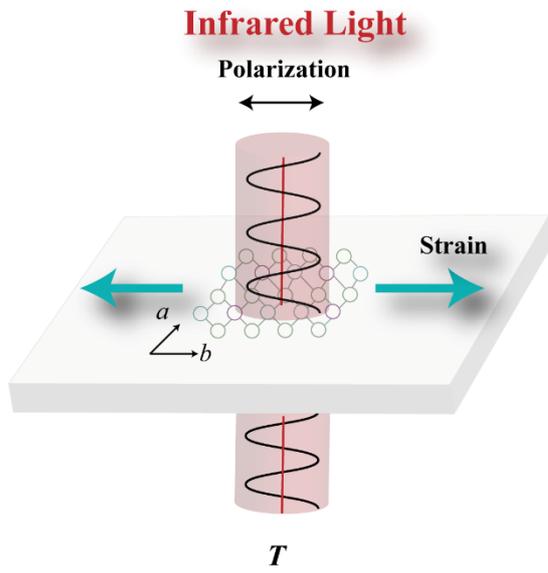
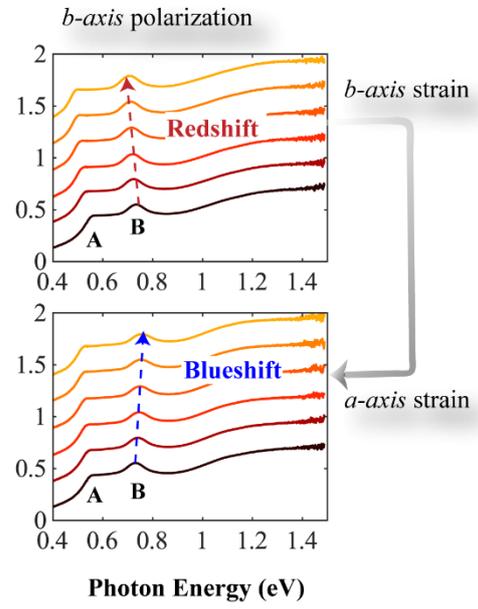